\PassOptionsToPackage{table}{xcolor}
\documentclass[conference]{IEEEtran}
\IEEEoverridecommandlockouts
\usepackage{cite}
\usepackage{amsmath,amssymb,amsfonts}
\usepackage{algorithmic}
\usepackage{graphicx}
\usepackage{textcomp}
\usepackage{hyperref}
\usepackage{amsmath}

\usepackage{algorithm}
\usepackage{subcaption}
\usepackage{tikz}
\usetikzlibrary{automata,arrows,arrows.meta,positioning,calc,patterns,decorations, decorations.markings,external,shapes.geometric}
\tikzexternaldisable
\usepackage{pgfplots}
\usepackage{pgfplotstable}
\usepackage{tabularx}
\usepackage{booktabs}
\usepackage[absolute,showboxes]{textpos}
\usepgfplotslibrary{groupplots,colormaps,colorbrewer,statistics,fillbetween}
\def\BibTeX{{\rm B\kern-.05em{\sc i\kern-.025em b}\kern-.08em
    T\kern-.1667em\lower.7ex\hbox{E}\kern-.125emX}}

\usepackage{pifont}

\usepackage{xspace}

\newcommand{\etal}{\textit{et al.\ }}

\newcommand{\ie}{\textit{i.e.,}~}

\newcommand{\one}{({\em i})\xspace}
\newcommand{\two}{({\em ii})\xspace}
\newcommand{\three}{({\em iii})\xspace}

\newcommand{\our}{$\text{CoRa}$\ }

\let\orgautoref\autoref
\renewcommand{\autoref}
{\def\sectionautorefname{Section}%
\def\subsectionautorefname{Section}%
\def\subsubsectionautorefname{Section}%
\orgautoref}

\makeatletter
\renewcommand{\paragraph}[1]{\vspace*{0.03in}\noindent{\bf #1.}\hspace{0.25ex \@plus1ex \@minus.2ex}}
\newcommand{\paragraphS}[1]{\vspace*{0.03in}\noindent{\bf #1}\hspace{0.25ex \@plus1ex \@minus.2ex}}
\makeatother

\begin{document}

\title{CoRa: A Collision-Resistant LoRa\\ Symbol Detector of Low Complexity\\
\thanks{This work was funded by the German Federal Ministry of Education and Research within the projects C-ray4edge and RESCUE-MATE.}
}

\author{\IEEEauthorblockN{Jos{\'e} \'Alamos}
\IEEEauthorblockA{
\textit{HAW Hamburg}\\
jose.alamos@haw-hamburg.de}
\and
\IEEEauthorblockN{Thomas C. Schmidt}
\IEEEauthorblockA{
\textit{HAW Hamburg}\\
t.schmidt@haw-hamburg.de}
\and
\IEEEauthorblockN{Matthias W{\"a}hlisch}
\IEEEauthorblockA{
\textit{TU Dresden and Barkhausen Institut}\\
m.waehlisch@tu-dresden.de}
}

\maketitle

\setlength{\TPHorizModule}{\textwidth}
\setlength{\TPVertModule}{\paperheight}
\TPMargin{5pt}
\begin{textblock}{1}(.1,0.01)
\noindent
\footnotesize
If you refer to this paper, please cite the peer-reviewed publication::
J. Alamos, T. C. Schmidt, M. W{\"a}hlisch.\\
CoRa: A Collision-Resistant LoRa Symbol Detector of Low Complexity.\\
\emph{Proc. of the 2025 IEEE INTERNATIONAL CONFERENCE ON COMPUTER COMMUNICATION
AND NETWORKS., 2025.}
\end{textblock}

\begin{abstract}
Long range communication with LoRa has become popular as it avoids the complexity
of multi-hop communication at low cost and low energy consumption. 
LoRa is openly accessible, but its packets are particularly vulnerable to
collisions due to long time on air in a shared band. This
degrades communication performance. Existing techniques for demodulating
LoRa symbols under collisions face challenges such as high computational
complexity, reliance on accurate symbol boundary information, or error-prone
peak detection methods.
In this paper, we introduce \our, a symbol detector for 
demodulating LoRa symbols under severe collisions.
\our employs a Bayesian classifier to accurately identify the true symbol amidst
interference  from other LoRa transmissions, leveraging empirically derived
features from raw symbol data. Evaluations using real-world
and simulated packet traces demonstrate that \our clearly outperforms the related state-of-the-art, \ie  up to 29\%  better decoding performance than TnB and 178\% better than CIC.  Compared to the LoRa baseline demodulator, \our magnifies the packet reception rate by up to $11.53\times$. \our offers a significant reduction
in computational complexity compared to existing solutions by only adding a constant overhead to the baseline demodulator,
while also eliminating the need for peak detection and accurately identifying colliding frames.
\end{abstract}


\section{Introduction}

Low Power Wide Area Network (LPWAN) technologies are emerging as a promising
solution for the Internet of Things (IoT), as these avoid the complexity of
multi-hop communication while maintaining low deployment costs.
The inherent characteristics of LPWAN technologies, \ie long-range
communication  and low power consumption, make them suitable for a
diverse range of IoT use cases, including environmental monitoring, asset tracking, 
and smart cities. Hence, LPWAN technologies are expected
to continuously attract significant interest and investment in the coming years.

 LoRa has emerged as a particularly
noteworthy solution in both industry and  academia due to its flexibility, wide  availability, and cost-effectiveness. 
 LoRa can operate in unlicensed spectrum, such as the 868 MHz band in Europe,
the 915 MHz band in the United States and the 2.4 GHz band worldwide.
It is typically used within the LoRaWAN architecture, which provides a cloud-based MAC layer
to facilitate device communication.
The LoRa modulation utilizes a variant of the chirp spread spectrum (CSS) technique, which
enables a broad transmission range of several kilometers in urban environments and 
tens of kilometers in rural areas. This is achieved through the use of a linearly
modulated chirp that spreads the transmission over a wide frequency band, resulting
in robustness against multipath fading, interference, and the Doppler effect.

Existing studies~\cite{hm-cltcs-21} estimate 42,000
devices per square kilometer in urban deployments, with each device transmitting
100-byte payloads, 300 times per day. In the EU868 region,
with 8-channel LoRaWAN gateways, assuming 35\% of device employ spreading factor 7 (bandwidth 125\,kHz),
a gateway covering an 800\,m radius would receive about 10 packets per second per channel. Given
a 190\,ms time-on-air (ToA) for each packet, no more than 5 packets per second can be transmitted
without overlap, making collisions unavoidable. While LoRa frames can be decoded if
the received signal is stronger than the colliding one (capture effect), collisions of similar power
 prevent decoding.

Recent work  has
studied the decoding of LoRa frames under collisions to increase the overall utility and 
bandwidth  of LoRa networks and could demonstrate a significant increase
in network capacity. However, these approaches often entail high computational complexity,
pose challenges for real-time operation or require a prior knowledge of the position
of interference frames relative to the target frame. The latter is ineffective
if the frame detection algorithm fails to detect interfering frames, a
situation that is common under severe frame collisions or a low Signal-to-Noise
Ratio (SNR).
These limitations challenge implementability on real-world devices.

In this work, we propose CoRa, an alternative symbol detector designed to
accurately classify LoRa symbols by employing a Bayesian classifier. \our is
specifically engineered to identify the true symbol within a frame amidst
interferences in the received signal, all while maintaining low computational complexity. This approach enhances the
demodulation accuracy of LoRa symbols in congested environments, leading to a
significant improvement in the Packet Reception Ratio (PRR). Our contributions read:

\begin{enumerate}
\item We design a new detection feature based on waveform symmetries, which arms our Bayesian classifier.
\item We demonstrate the ability to accurately demodulate symbols under
    collisions and under significant channel fluctuations by employing our robust symbol
        detector.
\item By leveraging the current state-of-the-art LoRa decoder (TnB Block Error Correction) in real-world captures,
    we achieve an improvement of Packet Reception Ratios (PRRs) of up to 29\% over the current state-of-the-art and approximately 11.53 times better performance than the baseline LoRa decoder.
\end{enumerate}

The remainder of this paper is structured as follows.
\autoref{sec:background} provides the necessary background on LoRa transmission along with related work.  \autoref{sec:decoder} details the proposed design for the LoRa
symbol detector. We present the evaluation on real-world and simulated data in
\autoref{sec:evaluation} and a discussion in
\autoref{sec:discussion}. Finally, we conclude the paper in
\autoref{sec:conclusions} with an outlook on future research.

\section{Background and Related Work}\label{sec:background}

A LoRa symbol $x(t)$ (\autoref{fig:lora_demodulation}, top) encodes information by shifting the frequency offset
of a linearly modulated complex chirp $x_0(t) = e^{j(\pi \frac{B}{T_s} t^2)}$ with bandwidth $B$ and symbol time $T_s = \frac{2^{SF}}{B}$,
where $SF$ denotes the spreading factor of the signal.
This process
transforms the symbol into the form $x(t) = x_0(t) \cdot e^{j 2 \pi f_m}$,
wherein $f_m$ corresponds to the frequency offset.

The LoRa signal is characterized by the signal circumvolution at the boundary
values, analogous to sampling at its Nyquist
frequency, $B$. This approach provides two benefits: \one
efficient modulation of LoRa symbols via cyclic shifts of the reference complex
chirp ($x_0$) and \two confinement of the entire signal within the
bandwidth $B$, conserving bandwidth.

In addition to the bandwidth and spreading factor,
LoRa modulation incorporates a coding rate parameter that
governs the inclusion of redundancy bits, enhancing error correction and
robustness under varying channel conditions.

\subsection{LoRa Demodulation}

\begin{figure}
\begin{tikzpicture}
\tikzstyle{every node}=[font=\small]

\begin{groupplot}[
    group style={
      group size=1 by 3,
      horizontal sep=3.5cm,
      vertical sep=1cm,
    },
    height=0.10\textheight,
    width=0.5\textwidth,
    xtick={0,10,20,30},
    xticklabels={},
    xmin=0, xmax=30,
    ymin=0, ymax=10,
    xlabel={},
    ylabel={},
    ylabel style={yshift=-0.5cm},
    ytick={0,10},
    yticklabels = {0,B},
    title style={yshift=0cm},
    ]
    \nextgroupplot[title=\textbf{$X$ (LoRa symbol)}]
    \addplot[thick, mark=none] coordinates {
        (0,4) (6,10)

        (6,0) (10,4)

        (10,1) (19,10)

        (19, 0) (20,1)

        (20,8) (22,10)

        (22,0) (30, 8)
    };

    \addplot[forget plot, dashed] coordinates {(10,0) (10,10)};
    \addplot[forget plot, dashed] coordinates {(20,0) (20,10)};
    \nextgroupplot[ylabel={Frequency}, title=\textbf{$\overline{X_0}$ (Downchirp)}]
    \addplot[thick, mark=none] coordinates {
        (0,10) (10,0)

        (10,10) (20,0)

        (20,10) (30,0)
    };

    \addplot[forget plot, dashed] coordinates {(10,0) (10,10)};
    \addplot[forget plot, dashed] coordinates {(20,0) (20,10)};

    \nextgroupplot[thick, xticklabels={0, $T_s$, $2T_s$, $3T_s$}, xlabel=Time, title=\textbf{$X_d$ (Dechirped symbol)}]
    \addplot[mark=none] coordinates {
        (0,4) (10,4)

        (10,1) (20,1)

        (20,8) (30,8)
    };

    \addplot[forget plot, dashed] coordinates {(10,0) (10,10)};
    \addplot[forget plot, dashed] coordinates {(20,0) (20,10)};
\end{groupplot}
\path (group c1r1.north west) -- (group c1r1.north east) node[pos=0.17, anchor=south,font=\tiny] {Symbol 1} node[anchor=south, pos=0.5, font=\tiny]{Symbol 2} node[pos=0.8333, anchor=south, font=\tiny]{Symbol 3};
\end{tikzpicture}
\caption{Spectrogram of a LoRa symbols (top), downchirp signal (middle) and dechirped symbols used for demodulation (bottom).}
\label{fig:lora_demodulation}
\end{figure}


To demodulate LoRa symbols, the receiver synchronizes the signal to the symbol
boundary and multiplies it with the complex conjugates of the reference chirp
signal (Downchirp signal, $\overline{x_0(t)}$, \autoref{fig:lora_demodulation},
middle). The resulting signal $X_d(t)$ (\autoref{fig:lora_demodulation},
bottom), referred to as the dechirped signal, is a complex waveform
characterized by frequency $f_m$, phase $\theta$ and magnitude $A$

$$
x_d(t) = \underbrace{\overline{x_0(t)} \cdot x_0(t)}_{= 1} \cdot A e^{j 2 \pi f_m + \theta} = A e^{j 2 \pi f_m + \theta}
$$

To recover the frequency offset encoding the symbol information, the
receiver performs a Fast Fourier Transform (FFT) on each dechirped symbol.
Sampling at rate $Fs=B$ results in $N = 2^{SF}$ frequency bins.
The symbol $m$ is determined by the bin with the highest magnitude.
Phase $\theta$ is discarded as it does not encode information.
The received signal typically includes additive white Gaussian noise
(AWGN), which distributes uniformly across bins, minimizing the likelihood
of noise misleading the demodulator.

\subsection{Synchronization and Offset Estimation}

The receiver detects a LoRa signal by identifying the preamble,
consisting of consecutive reference chirps ($x_0$), followed by two sync-word symbols
and two and a quarter downchirps ($\overline{x_0(t)}$). 
The receiver detects the presence of a frame upon identifying consecutive peaks
from the preamble and uses the downchirp symbols with
the preamble to correct the carrier frequency offset (CFO) and
the sample time offset (STO).

The CFO, due to minor discrepancies between transmitter and receiver oscillator
frequencies, and STO, the deviation from the actual symbol boundary, affect the
frequency of the dechirped signal and FFT bin accuracy. The integer part of CFO
and STO can be determined using the preamble and downchirp, but fractional
components are challenging.
Fractional offsets spread target bin energy across adjacent bins, complicating
symbol detection in low SNR scenarios. Oversampling and fractional offset
estimation techniques, such as those by Berner et al.~\cite{bdd-lclfs-20}, help
estimate these components but do not fully resolve the issue.

\subsection{Effect of Collisions}\label{sec:background_collisions}

In a two-packet collision, we analyze
a symbol with a frequency offset $f_m$. The symbol boundary offset $\Delta t$ marks
the start of the interfering symbol relative to the original 
symbol. This results in three waveforms in the
dechirped signal: a complete waveform at frequency $f_m$, a clipped waveform
spanning from 0 and $\Delta t$ at frequency $f_1$ and another clipped waveform
spanning from $\Delta t$ and $T_s$ at frequency $f_2$. These phenomena
are illustrated in \autoref{fig:lora_collisions} (left).

The spectrum of the dechirped symbol (\autoref{fig:lora_collisions}, right)
exhibits three peaks, corresponding to the three waveforms. In cases where the colliding symbols
pose comparable energy levels, the distribution of the interference symbol energy across the
peaks at $f_1$ and $f_2$ typically preserves the target peak as the most prominent
among all observed peaks. However, in scenarios where the energy of the colliding symbols exceeds the
energy of the true symbol, the peaks at $f_1$ and $f_2$ may display greater energy than the
true symbol, which leads to an incorrect classification of the symbol. This particular scenario
is illustrated in \autoref{fig:lora_collisions} (right).

\begin{figure}
\begin{tikzpicture}
\tikzstyle{every node}=[font=\small]
\begin{groupplot}[
group style={
  group size=2 by 1,
  horizontal sep=1.1cm,
  vertical sep=0.50cm,
},
height=0.23\textwidth,
width=0.25\textwidth,
axis on top,
yticklabel pos=left,
xmin=0, ymin=0,
legend cell align={left},
legend style={
  fill opacity=0.8,
  draw opacity=1,
  text opacity=1,
  at={(0.05,0.99)},
  nodes={scale=0.8, transform shape},
  anchor=north west,
  draw=white!80!black,
},
tick align=outside,
tick pos=left,
x grid style={white!69.0196078431373!black},
xtick style={color=black},
y grid style={white!69.0196078431373!black},
ytick style={color=black},
cycle list name=exotic,
xtick = {10,18,20},
yticklabels = {},
xticklabels = {0,$\Delta t$,$T_s$},
ylabel style={
    yshift=-0.5cm,
},
]
\nextgroupplot[title=\textbf{Spectrogram}, ylabel={Frequency},xlabel={Time},xmin=10,xmax=20, ymax=10,
ytick={0,1,4,8,10},
yticklabels = {0,$f_2$,$f_m$,$f_1$,B},
    ]
    \addplot[thick,mark=none] coordinates {
        (0,6) (10,6)

        (10,4) (20,4)

        (20,8) (30,8)
    };
    \addplot[thick,mark=none] coordinates {
        (-2,5) (8,5)

        (8,8) (18,8)

        (18,1) (28,1)
    };

    \nextgroupplot[title={\textbf{Frequency spectrum}}, ylabel={Magnitude},xlabel={Frequency},xmin=0,xmax=128, xtick={0,128}, xticklabels={0,B},
ylabel style={yshift=-0.2cm},
    xtick={0,0.1*128,0.5*128,0.8*128,128},
    xticklabels = {0,$f_2$,$f_m$,$f_1$,B},
    ytick={\pgfkeysvalueof{/pgfplots/ymin}, \pgfkeysvalueof{/pgfplots/ymax}},
    yticklabels = {0,$A_{max}$},
    ]

    \addplot[only marks, mark=*, mark size=2, mark options={fill=green}] coordinates {(64,127.89)};
    \addlegendentry{True peak}
    \addplot[only marks, mark=*, mark size=2, mark options={fill=red}] coordinates {(102,202.08)};
    \addlegendentry{Detected}
    \addplot[thick, mark=none,thick] table [x expr=\coordindex, y index=0] {data/lora_symbol_fft.csv};
\end{groupplot}
\end{tikzpicture}
\caption{Illustration of a LoRa symbol subjected to collision, showing both the spectrogram (left) and frequency spectrum (right). The baseline LoRa demodulator selects the peak at $f_1$, in contrast to the true peak location in $f_m$.}
\label{fig:lora_collisions}
\end{figure}



\subsection{Related Work}\label{sec:related_work}

LoRa transmissions are highly susceptible to collisions~\cite{f-cplal-17,ofjg-cccal-19}.
Several authors propose countermeasures to this problem in the context of
LoRaWAN networks. Rizzi~\etal~\cite{rffsg-uliwn-17} and Leonardi ~\etal~\cite{lbbp-calma-20}
propose light modifications to the LoRaWAN Media Access Control, which already
improve performance metrics of uplink-oriented Class-A deployments. Xu~\etal~\cite{xlyhd-sahsas-20}
introduce an adaptive scheduling mechanism for LoRaWAN networks, that organizes
devices based on their transmission patterns, with the purpose of reducing packet
collisions between groups. Vincenzo~\etal~\cite{vht-idsl-19}
improves the performance of downlink traffic for scenarios with low to medium
downlink load, by improving gateway selection for downlink traffic and extending
the number of gateways --- at the price of increased deployment
costs.

Existing work addresses the inherent constraints of LoRa transmissions through
the employ of Media Access Control (MAC) strategies. Several authors propose
to employ multi-hop LoRa networks to improve energy efficiency and overall
communication performance~\cite{ck-lmnrc-20,stbbs-ermcb-17,tbs-eticl-17,aa-mllbm-19,bgbts-teelm-19}.
There exists a growing interest in Time-Division Multiple Access schemes for LoRa
communication. Zorbas \etal~\cite{zakp-ttlii-20} propose a time-slotted scheme
for LoRa traffic, which focus on applications that require frequent and very
reliable communication. Haubro \etal~\cite{hoof-tlrri-20} and Alamos
\etal~\cite{aksw-dslrc-22} design adaptions layer to facilitate the operation of
TSCH and DSME modes, respectively, of IEEE 802.15.4 over the LoRa Physical
Layer (PHY). Due to the nature of time-slotted communication, these strategies
trade-off transmission delay with communication reliability. DSME-LoRa also enabled 6LoRa~\cite{asw-6fsin-23}, a full-featured IPv6 network integration into the RIOT operating system~\cite{bghkl-rosos-18}.

Multiple
studies~\cite{glgtl-lecmal-20,p-iecca-18,knwm-pecsl-20, lbb-nacal-23} have
examined contention-based communication in LoRa networks, targeting the
facilitation of unsynchronized communication, which is a common scheme in LoRaWAN
networks.

New Physical Layer (PHY) techniques have been proposed to extend the transmission
range of LoRa and to enable concurrent transmissions. Li \etal~\cite{lgtzc-ntusl-21} and Tesfay~\etal~\cite{tskc-dlsdu-21}
propose deep-learning based symbol detection to decode packets under very low
SNR conditions, effectively enhancing transmission range. Over recent years,
a comprehensive analysis has been carried out on multiple strategies for packet
transmission within environments vulnerable to packet collisions.
Eletreby \etal~\cite{ezky-elpwa-17} propose to unravel
packet collisions in LoRaWAN by exploiting the inherent differences in carrier frequency
offsets created by distinct devices. These
frequency variations arise due to imperfections in crystal oscillators,
a common characteristic of low-cost LoRa devices. FTrack~\cite{xzg-fpdlt-19} leverages both the time and
frequency domain to identify the continuity of the chirp signal and eliminate interference
symbols. The approach increases up to three times the network throughput at the cost of high
computational complexity. 

Shahid~\etal~\cite{spcbk-cicdm-21}~introduce the Concurrent Interference Cancellation (CIC) method that
utilizes symbol boundaries to delineate sub-symbols and cancel out interference symbols by
intersecting the sub-symbols spectrum. The algorithm incorporates a peak discrimination
strategy, where interference peaks from incomplete cancellation are filtered
out if their magnitude differs from the expected peak magnitude, which is derived from the
preamble peaks.
The algorithm complexity is driven by the need for the one Fast Fourier Transform (FFT)
to process the dechirped symbol, in addition to two additional FFTs for each colliding packets to analyze the sub-symbols. This
results in a computational complexity of $O(N log(N) (1 + 2 K)))$, where $N$ represents the number of data samples per symbol and $K$
denotes the number of colliding packets.
The authors have made available the original datasets along with both Matlab and Python implementations
of the demodulator. Nonetheless, discrepancies arise as the implementations yield divergent outcomes
when applied to the original datasets. Furthermore, the throughput results presented in the original study
could not be reproduced using these tools. The Paralign~\cite{wzzlm-dlcpa-23} method conducts spectral intersection
on transposed overlapping symbols to identify the actual symbols. Although this approach
offers computational complexity benefits compared to previous methods, its
effectiveness depends on the accurate detection of packets --- a process known to be challenging
in high-traffic conditions.

Tong~\etal~\cite{txw-cempr-20} propose CoLoRa, a method to locate peaks in a dechirped
symbol even under low SNR conditions and an algorithm to group symbols by peak power ratio,
in order to disentangle packet collisions. Rhati~\etal~\cite{rz-trclb-22}
propose the TnB LoRa demodulator, which incorporates a robust frame synchronization
algorithm, a symbol detector (Thrive), and a Block Error Correction (BEC)
decoder algorithm. The Thrive symbol detector leverages the observation that
true peaks appear as offset peaks within interference symbols,
identified as siblings. Thrive calculates a matching cost to assign each
demodulation window its sibling with the highest magnitude, corresponding
to the fully aligned symbol. The algorithm also includes a peak history
estimation to account for channel fluctuations. Unlike the Thrive
algorithm, our demodulator does not require peak detection or depend on
symbol boundary information. The BEC algorithm is designed to decode frames
containing one or more symbol errors, significantly enhancing the decoding
performance. We integrate the BEC algorithm in our work to ensure robust decoding
of LoRa frames.

\section{\our Design} \label{sec:decoder}

\begin{figure}
\centering
\includegraphics[width=0.5\textwidth]{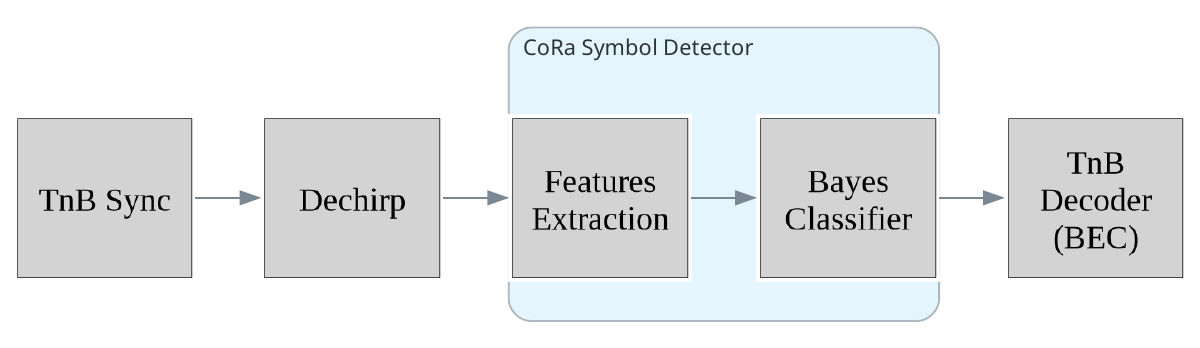}
\caption{System architecture of \our. The blue box highlights the symbol detector block, which is our main contribution}
\label{fig:system_architecture}
\end{figure}

We design a symbol detector that employs a Bayesian classifier to
estimate the value of a received symbol based on features computed directly from the
dechirped symbol.  This symbol detector identifies the true peak amidst interference peaks.
\autoref{fig:system_architecture} illustrates our system architecture, which utilizes components of TnB \cite{rz-trclb-22} to leverage the
synchronization and decoder blocks, but replaces  the Thrive detector of TnB.

The synchronization system remains in a state of readiness for incoming frames.
Upon detection of a frame, it estimates the sampling time offset and
carrier frequency offset. Following the standard LoRa
demodulation procedure, the system then dechirps the symbol by multiplying it with
the conjugate of the base upchirp. Subsequently, the symbol detector block
extracts pertinent features from the dechirped symbol, which then serve as input
to the Bayesian classifier. The Bayesian classifier proceeds to select the
symbol value with the highest likelihood. Following demodulation, the BEC
decoder (TnB) decodes the frames. Similar to existing solutions such as CIC or
TnB, the system is capable of concurrent operation.

\begin{figure}
\centering
\includegraphics[width=0.5\textwidth]{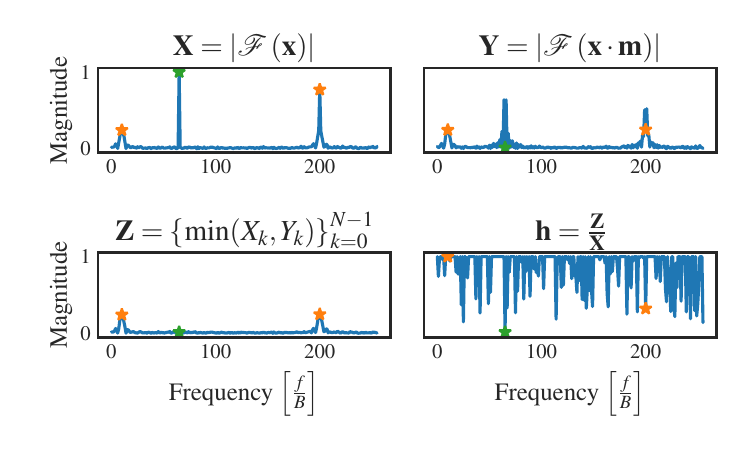}
\caption{Derivation of the Half-Period Discriminator (HPD), showing the spectrum of a dechirped symbol (X), the spectrum
    of the signal with a phase inversion in the second half (Y), the masked phase-inverted signal (Z) and the output of the HPD feature.
    The figure shows the location of the interference peaks (orange markers) as well as the true peak (green marker)}
\label{fig:hpd_derivation}
\end{figure}

\subsection{Feature Extraction}

We calculate two key features. The Peak Magnitude Deviation (PMD), which measures the discrepancy from the 
expected peak magnitude, and the Half-Period Discriminator (HPD), which exploits
wave symmetry properties and is a novel contribution of this work.

\paragraph{Peak Magnitude Deviation (PMD)}
The Peak Magnitude Deviation (PMD) quantifies the difference between the observed
FFT bin magnitude of a dechirped symbol and its expected peak
magnitude. While CIC~\cite{spcbk-cicdm-21} introduces a peak filtering method based on magnitude alignment,
our work builds on this principle by developing PMD as a distinct feature, applying it as a cost function to enhance
peak identification under interference. We introduce the PMD, denoted as $p_k$, by calculating the normalized deviation
of each $X_k$ from the expected peak magnitude, capping the result at 1 to avoid extreme values.

$$
\mathbf{p} = \left\{\min\left(\left|\frac{|X_k| - E(|X_{p}|)}{E(X_{p})}\right|, 1\right)\right\}_{k=0}^{N-1}
$$ 

\noindent where $X_k$ is the dechirped symbol spectrum at bin $k$ and $E(X_p)$
is the expected peak magnitude. $E(|X_p|)$ is estimated by averaging preamble
peak magnitudes during synchronization.

The PMD helps assess the likelihood of a bin being the true symbol.
The FFT bin containing the true peak typically aligns with the preamble peak
magnitude, resulting in near-zero PMD values.
Interference peaks typically produce PMD values that deviate from zero due to
misalignment and clipping. Similarly, noise bins often result in PMD deviations
from zero as their magnitudes are generally lower than the preamble. 

While PMD serves as a useful cost function, it can misrepresent the true peak in specific scenarios:
\one At low SNR scenarios, noise can distort the true symbol energy, causing deviations that increase
PMD and incorrectly penalize the true peak. \two When interference peaks match the preamble energy,
PMD values approach zero, failing to penalize interference. To address these limitations, we introduce a second feature -- the Half-Period Discriminator (HPD).

\paragraph{Half-Period Discriminator (HPD)}
The Half-Period Discriminator assesses the presence of symmetric or
antisymmetric periodic frequency components within a signal.

This metric is built upon the observation that a true peak manifests as a complete
waveform, exhibiting either symmetry (for even frequencies) or antisymmetry (for odd frequencies).
In contrast, interference peaks typically appear as clipped waveforms that do not conform
to symmetry patterns. This distinction aids in separating the
true peak from interference peaks.

Consider a discrete signal $\mathbf{x} = \{x_n\}_{n=0}^{N-1}$, and a signal \linebreak $m=\{m_n\}_{n=0}^{N-1}$ defined as 1 for $n<\frac{N}{2}$ and
-1 for $n \geq \frac{N}{2}$. 
The signal $\mathbf{y} = \{y_n\}_{n=0}^{N-1} = \{x_n \cdot m_n\}_{n=0}^{N-1}$ then represents the discrete signal $\mathbf{x}$ with a phase inversion in the second half of the signal.
Consider the Discrete Fourier Transform of $\mathbf{y}$, $Y=\{Y_k\}_{k=0}^{N-1}$:

\begin{align}
	Y_k &= \sum_{n=0}^{N-1} x_n m_n \cdot e^{-j 2 \pi \frac{k}{N}n} \nonumber\\
	&= \sum_{n=0}^{\frac{N}{2} - 1} x_n e^{-j 2 \pi \frac{k}{N}n} - \sum_{n=\frac{N}{2}}^{N-1} x_n e^{-j 2 \pi \frac{k}{N}n} \nonumber \\
	&= \sum_{n=0}^{\frac{N}{2} - 1} \left(x_n - (-1)^k x_{n + \frac{N}{2}}\right) \cdot e^{-j 2 \pi \frac{k}{N}n}\label{eqn:fourier}
\end{align}

\autoref{eqn:fourier} shows that the symmetry of a complete waveform $e^{j 2 \pi \frac{k}{N}n}$ yields $Y_k = 0$. In contrast,
the peak of a clipped waveform attains a non-zero spectrum, as the
clipped signal is neither symmetric nor antisymmetric.
\autoref{fig:hpd_derivation} illustrates $\mathbf{X} = \left|\mathcal{F}(\mathbf{x})\right|$, the magnitude of the spectrum of a dechirped symbol under collisions (top left), along with $\mathbf{Y}$, the magnitude of the spectrum after phase inversion (top-right).

Notable, although the true peak in $\mathbf{Y}$ exhibits energy in its neighboring components,
the true peak itself yields zero, whereas other peaks show values greater than
zero. We apply the minimum between the magnitude of $\mathbf{X}$ and $\mathbf{Y}$ to mask
out the energy components around the true peak, resulting in the signal $\mathbf{Z} = \{ Z_k \}_{k=0}^{N-1}=\{\min(X_k,Y_k)\}_{k=0}^{N-1}$ (bottom-left). This signal
retains energy at the interference peaks, while the values at the true peak the value remain low.
We construct the HPD, denoted as $h_k$, by dividing $\mathbf{Z}$ by $\mathbf{X}$, \ie normalizing the feature to the range [0, 1].

$$
\mathbf{h} = \frac{\mathbf{Z}}{\mathbf{X}} = \left\{\frac{\min(|X_k|, |Y_k|)}{|X_k|}\right\}_{k=0}^{N-1}
$$

As illustrated in \autoref{fig:hpd_derivation} (bottom-right), the HPD values
at the peaks align with our previous analysis.

However, similar to PMD, HPD alone is not sufficient to reliably identify the true peak in all scenarios:
\one Noise may exhibit symmetric or antisymmetric components, resulting in HPD values close to zero.
This is shown in \autoref{fig:hpd_derivation} (bottom-left).
\two While carrier frequency offset (CFO) and sample time offset (STO) can be estimated and cancelled,
poor estimation can cause the true peak to appear at a fractional frequency, reducing the effectiveness of phase cancellation and leading to higher HPD values.

The challenges associated with PMD and HPD are distinct and orthogonal. This motivates the development of a robust
estimator that combines both features through a Bayesian Classifier to improve peak discrimination.

\subsection{Bayesian Classifier}\label{sec:bayesian}

Consider a dechirped symbol $x = \{x_n\}_{n=0}^{N-1}$ obtained from a signal captured at a
LoRa receiver. The spectrum of this signal exhibits three types of peaks: \one
the true peak, characterized by a complete waveform with a magnitude similar
to the preamble. \two, interference peaks from data symbols, characterized by
clipped waveforms with varying magnitude; and \three interference peaks from
preamble symbols, characterized by two clipped waveforms with identical
frequencies, which resemble a complete waveform.
Since interference preamble peaks are seen as complete waveforms and may
exhibit similar energy to the true peak, they can resemble the true peak.
Therefore, detecting a complete waveform is insufficient for correct symbol
detection. However, the position of the true peak is likely to vary across
consecutive symbols due to the whitening component of the LoRa modulator, which
scrambles the data and makes the data peak position independent of the previous
symbol. In contrast, preamble peaks consistently arise in the same position
across consecutive symbols.

Let $T_{k}\left(\mathbf{x}\right)$ and $C_{k}\left(\mathbf{x}\right)$ denotes the events that the $k$ of the spectrum
of the dechirped symbol $x$ contain the true peak and the expected complete
waveform, respectively. Let $p_{k}\left(\mathbf{x}\right)$ and $h_{k}\left(\mathbf{x}\right)$ be the values of the PMD 
and the HPD, respectively, of
symbol $\mathbf{x}$ at bin $k$. Let \( \mathbf{ph}_{k} \left(\mathbf{x}\right) = \left(p_{k}\left(\mathbf{x}\right), h_{k}\left(\mathbf{x}\right)\right) \). Leveraging
the independence between values of consecutive data symbols, we calculate
the Bayesian classifier $P\left(T_{k}\left(\mathbf{x}\right)\right)$ as the probability of containing the true peak at $k$,
considering $\mathbf{x}$ as the current symbol and $\mathbf{x}'$ as the previous symbol.
Specifically, $T_{k}\left(\mathbf{x}\right)$ reflects the process whereby the symbol $\mathbf{x}$ has the true peak at $k$ and true peak in the
previous symbol appears at a different location, thereby ensuring that the peak is not a preamble peak.

\begin{align}
    P\left(T_{k}\left(\mathbf{x}\right)\right) 
    &= P\left(C_{k}\left(\mathbf{x}\right) \mid \mathbf{ph}_{k}\left(\mathbf{x}\right) \cap \neg C_{k}\left(\mathbf{x}'\right) \mid \mathbf{ph}_{k}\left(\mathbf{x}'\right)\right) \nonumber \\
    &= P\left(C_{k}\left(\mathbf{x}\right) \mid \mathbf{ph}_{k}\left(\mathbf{x}\right)\right) \nonumber \\
    &\quad \cdot \left(1 - P\left(C_{k}\left(\mathbf{x}'\right) \mid \mathbf{ph}_{k}\left(\mathbf{x}'\right)\right)\right)
    \label{eq:bayes}
\end{align}

The true peak of symbol $\mathbf{x}$ is identified by selecting the bin $k$ that yields the highest probability.

Utilizing Bayes's theorem, the posterior probability
$P\left(C_{k} \mid \mathbf{ph}_{k}\left(\mathbf{x}\right)\right)$ can be estimated from the prior probability as follows:

\begin{equation}
    P\left(C_{k}\left(\mathbf{x}\right) \mid \mathbf{ph}_{k}\left(\mathbf{x}\right)\right) = \frac{P\left(\mathbf{ph}_{k}\left(\mathbf{x}\right) \mid C_{k}\left(\mathbf{x}\right)\right) \cdot P\left(C_{k}\left(\mathbf{x}\right)\right)}{P\left(\mathbf{ph}_{k}\left(\mathbf{x}\right)\right)}
    \label{eq:posterior}
\end{equation}

This equation indicates that by knowing the probabilities
$P\left(\mathbf{ph}_{k}\left(\mathbf{x}\right) \mid
C_{k}\left(\mathbf{x}\right)\right)$, $P\left(C_{k}\left(\mathbf{x}\right)\right)$ and
$P\left(\mathbf{ph}_{k}\left(\mathbf{x}\right)\right)$, we can calculate the probability of the
true peak using the Bayes classifier (see \autoref{eq:bayes}).

\subsection{Parameter Estimation}

The posterior probability (see \autoref{eq:posterior}), necessary for the
Bayesian classifier (see \autoref{eq:bayes}), can be determined if 
\( P\left(\mathbf{ph}_{k}\left(\mathbf{x}\right) \mid C_{k}\left(\mathbf{x}\right)\right) \), 
\( P\left(C_{k}\left(\mathbf{x}\right)\right) \), and 
\( P\left(\mathbf{ph}_{k}\left(\mathbf{x}\right)\right) \) 
are known. According to the law of total probability:

\begin{align}
    P\left(\mathbf{ph}_{k}(\mathbf{x})\right) &= P\left(\mathbf{ph}_{k}(\mathbf{x}) \mid C_{k}(\mathbf{x})\right) \cdot P\left(C_{k}(\mathbf{x})\right) \nonumber \\
                   &\quad + P\left(\mathbf{ph}_{k}(\mathbf{x}) \mid \neg C_{k}(\mathbf{x})\right) \cdot P\left(\neg C_{k}(\mathbf{x})\right)
\label{eq:total_prob}
\end{align}

\noindent This reduces the problem to determining \( P\left(\mathbf{ph}_{k}(\mathbf{x}) \mid
C_{k}(\mathbf{x})\right) \), \( P\left(\mathbf{ph}_{k}(\mathbf{x}) \mid \neg
C_{k}(\mathbf{x})\right) \), and \( P\left(C_{k}(\mathbf{x})\right) \).

To estimate $P\left(\mathbf{ph}_{k}\left(\mathbf{x}\right) \mid C_{k}\right)$ and
$P\left(\mathbf{ph}_{k}\left(\mathbf{x}\right) \mid \neg C_{k}\right)$, we simulate samples of $\mathbf{ph}_{k}$ at both the true peak
and the interference or noise peaks.
We then compute smoothed 2D histograms from these samples to calculate
the discrete probabilities.

The probability $P\left(C_{k}\right)$ is derived from the ratio of the true peak samples to
the total number of samples.

To create the true peak and interference/noise sample sets, we extract $\mathbf{ph}_{k}$ samples from 100,000 dechirped
symbols, each containing 256 bins. Each symbol comprises
a complete waveform (which represent the true peak) and up to two interference
packets, each characterized by two clipped waveforms, which are uniformly
distributed in terms of their frequencies and positions.
Transmission power is uniformly
distributed between -15 dB and 13 dB, To account for synchronization errors, a
small fractional frequency deviation, uniformly distributed within $\pm\frac{1}{8}$ of the
Nyquist frequency, is added to the symbol frequency.

For robust estimation, we focus on challenging symbols
misclassified by the baseline LoRa demodulator, for which the true peak
magnitude is not the most prominent. The true peak set comprises the $p_k$ and $h_k$
features of the bin containing the true peak. The interference set includes the 10 samples
of $\mathbf{ph}_{k}$ feature pairs with the lowest Peak Magnitude Deviation ($p_k$), ensuring the
Bayesian classifier emphasizes the HPD feature ($h_k$) for accurate classification.

\begin{figure}
\centering
\includegraphics[width=0.3\textwidth]{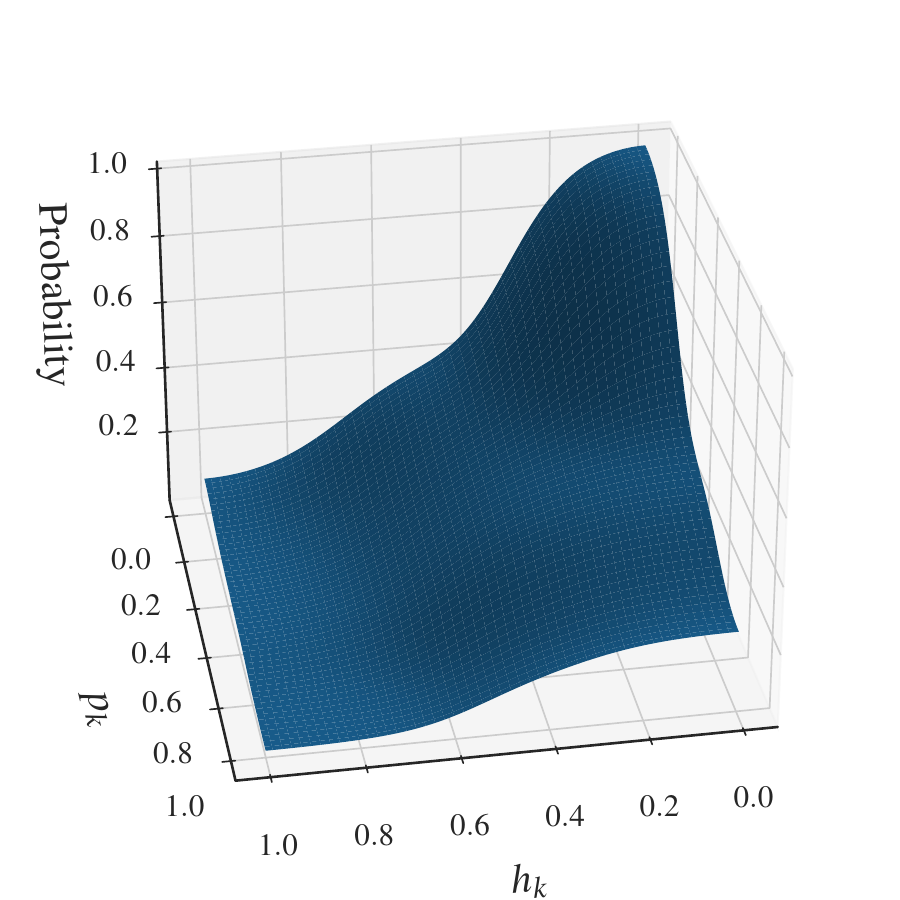}
    \caption{Discrete (200x200 samples) posterior probability $P\left[C_{k}\left(\mathbf{x}\right) \mid
    \mathbf{ph}_{k}\left(\mathbf{x}\right)\right]$, as a function of PMD ($p_k$) and HPD ($h_k$) features, estimated
    from simulated data. This
    probability is subsequently used to compute the Bayesian classifier $P\left(T_{k}\left(\mathbf{x}\right)\right)$.}
\label{fig:estimation}
\end{figure}

By using the expressions for the discrete probabilities and the total probability,
we obtain the discrete probability $P\left[C_{k} \mid \mathbf{ph}_{k}\right]$ depicted in
\autoref{fig:estimation}.
As anticipated, low values of $p_k$ and $h_k$, which characterize the true peak,
yield high probabilities. Conversely, high values of $p_k$ and $h_k$, which
are indicative of noise or interference peaks, result in low
probabilities. 

The posterior probability can be evaluated using interpolation techniques such
as nearest neighbor or linear interpolation. For memory constrained devices, function
approximations such as spline interpolation are feasible.

Using the evaluated posterior probability in \autoref{eq:bayes}, we derive the Bayes classifier.

\section{Evaluation}\label{sec:evaluation}
We now evaluate the performance
of our proposed demodulator in comparison to the standard LoRa receiver (baseline), and the two state-of-the-art demodulators CIC and TnB.
We utilize the CIC demodulator and the baseline
implementation from OpenLoRa~\cite{mkskc-ovlit-23}. For TnB, we use the reference MATLAB implementation~\cite{rz-trclb-22}.
Our demodulator is implemented in \textit{Python} using the \textit{NumPy}
library. To integrate the synchronization and decoding blocks of TnB, we
interface our solution with TnB by
employing the MATLAB Engine API \cite{matlab-engine-python}.
The posterior probability (see \autoref{fig:estimation}),
used to evaluate the Bayesian classifier, is
implemented using nearest neighbor interpolation with a grid
of 200x200 samples. This grid maps the calculated HPD and
PMD values to the closest value in the discrete probability estimated from
simulated data. This method ensures the accuracy of the classifier while
maintaining computational efficiency. We conduct the evaluation on a
\textit{Quad-Core Intel Core i7-8650U} Linux machine running at 1.9GHz with
16\,GB of RAM.

To foster a fair comparison, we conduct the analysis on the two datasets provisioned by the related work. Both datasets contain real-world LoRa signals of fixed transmitter devices.
The first dataset (from CIC) includes four scenarios with varying SNR conditions and fixed PHY parameters.
The second dataset (from TnB) comprises three scenarios with varying SNR conditions and varying PHY parameters.
The characteristics of these datasets, as well as the transmission parameters, are summarized in \autoref{tab:tx_params}.
All captures involve 20 transmitter devices, each sending a 12-byte payload.

\begin{table*}
    \centering
    \rowcolors{2}{gray!15}{white}
    \begin{tabularx}{\textwidth}{X|cccc|ccc}
	    \rowcolor{white}
        \multicolumn{1}{c}{} & \multicolumn{4}{c}{\textbf{CIC Dataset}} & \multicolumn{3}{c}{\textbf{TnB Dataset}} \\ 
        \cmidrule(lr){2-5} \cmidrule(lr){6-8}
        \textbf{Parameter} & \textbf{D1} & \textbf{D2} & \textbf{D3} & \textbf{D4} & \textbf{Indoor} & \textbf{Outdoor1} & \textbf{Outdoor2} \\ 
    \midrule
        Spreading Factor & \multicolumn{4}{c|}{8} & \multicolumn{3}{c}{8, 10} \\ 
        Bandwidth [kHz] & \multicolumn{4}{c|}{250} & \multicolumn{3}{c}{125} \\ 
        Coding Rate & \multicolumn{4}{c|}{4/5} & \multicolumn{3}{c}{4/5, 4/6, 4/7, 4/8} \\ 
        Sampling Rate [MHz] & \multicolumn{4}{c|}{2} & \multicolumn{3}{c}{1} \\ 
        SNR (SF8) [dB] & 30 to 42 & 30 to 42 & 10 to 30 & -17 to 5 & -5 to 17 & -10 to 10 & -8 to 14 \\ 
        SNR (SF10) [dB] & \multicolumn{4}{c|}{-} & -9 to 23 & -17 to 12 & -13 to 12 \\ 
	Deployment area [m$^2$] & $15 \times 10$ & $100 \times 60$ & $170 \times 100$ & $1200 \times 1600$ & $100 \times 120$ & $400 \times 240$ & $220  \times  200$ \\ 
        TX Rate [pkt/s] & \multicolumn{4}{c|}{5 to 100} & \multicolumn{3}{c}{20, 25} \\ 
        Capture Time [s] & \multicolumn{4}{c|}{60} & \multicolumn{3}{c}{30} \\ 
    \bottomrule
    \end{tabularx}
    \caption{Configuration parameters of LoRa devices across captured scenarios}
    \label{tab:tx_params}
\end{table*}

\subsection{Collision Resolution}\label{subsec:detection-performance}

\begin{figure}
    \includegraphics[width=0.5\textwidth]{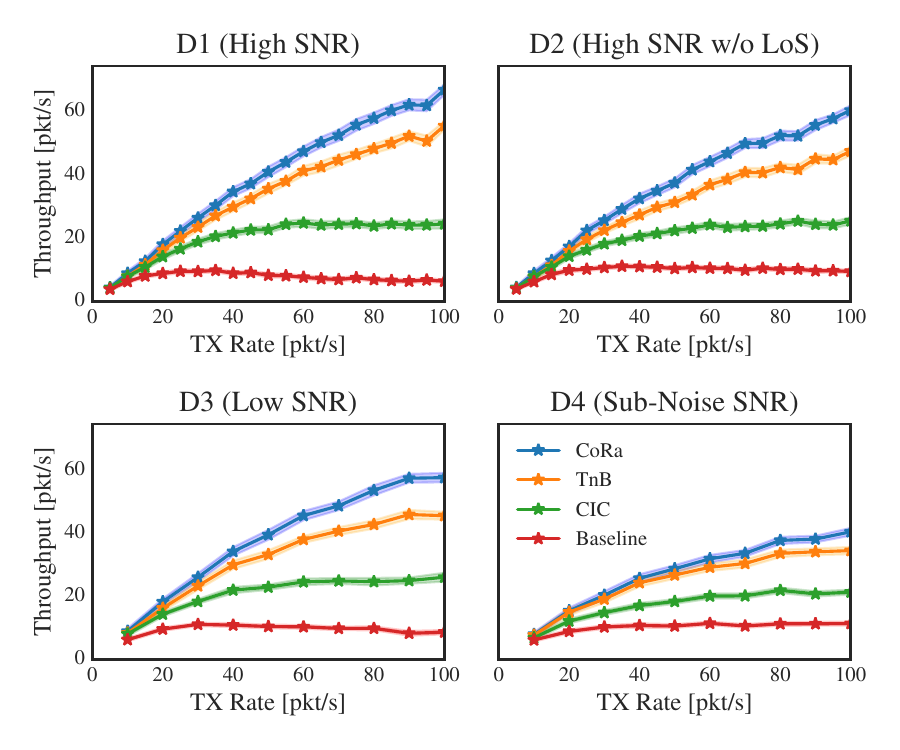}
    \caption{Throughput vs aggregated transmission rate in the CIC dataset. The shaded areas around the curves indicate the error margins.}
\label{fig:cic_dataset}
\end{figure}

\autoref{fig:cic_dataset} depicts the throughput of the baseline, CIC, TnB, and the \our demodulator using the CIC dataset.
 The \our approach demonstrates superior
throughput in all scenarios, achieving up to 29\%  improvement over TnB in high SNR conditions without Line-of-Sight (see \autoref{fig:cic_dataset}, D2 at 95 pkt/s)
and up to a 6\% improvement in the Sub-Noise SNR scenario (see \autoref{fig:cic_dataset}, D4 at 100 pkt/s).
Given that our demodulator utilizes the synchronization and decoding blocks of TnB, the observed
performance difference is attributed to the superior efficacy of our demodulator over the Thrive
demodulator utilized in TnB. Correspondingly, our solutions exhibits up to a 178\% improvement over
CIC and delivers up to 11.53 times higher throughput than the baseline in high SNR conditions (see \autoref{fig:cic_dataset}, left, 100 pkt/s).

Consistently,  TnB shows better results than CIC for two main reasons.
\one the effectiveness of CIC decreases with frame collisions, in which symbol boundaries
are closer than 10\% of the symbol time -- a situation that occurs in approximately 20\% of the collisions.
\two the correction capabilities of the BEC scheme utilized by TnB is superior to those of the
standard LoRa decoding scheme.

For the baseline, performance deteriorates significantly with an increased
aggregated transmission ratio and a decrease in Signal-to-Noise Ratio (SNR).
This is mainly due to the synchronization procedure of the standard demodulator, which
identifies the preamble by detecting consecutive symbols of identical value.
Frequent frame collisions in combination with low SNR can corrupt the demodulated
values, thereby hindering accurate frame detection. In contrast, the frame
detection methods of CIC and TnB remain effective even under frame collisions and 
 increase their throughput.

\begin{figure}
    \includegraphics[width=0.5\textwidth]{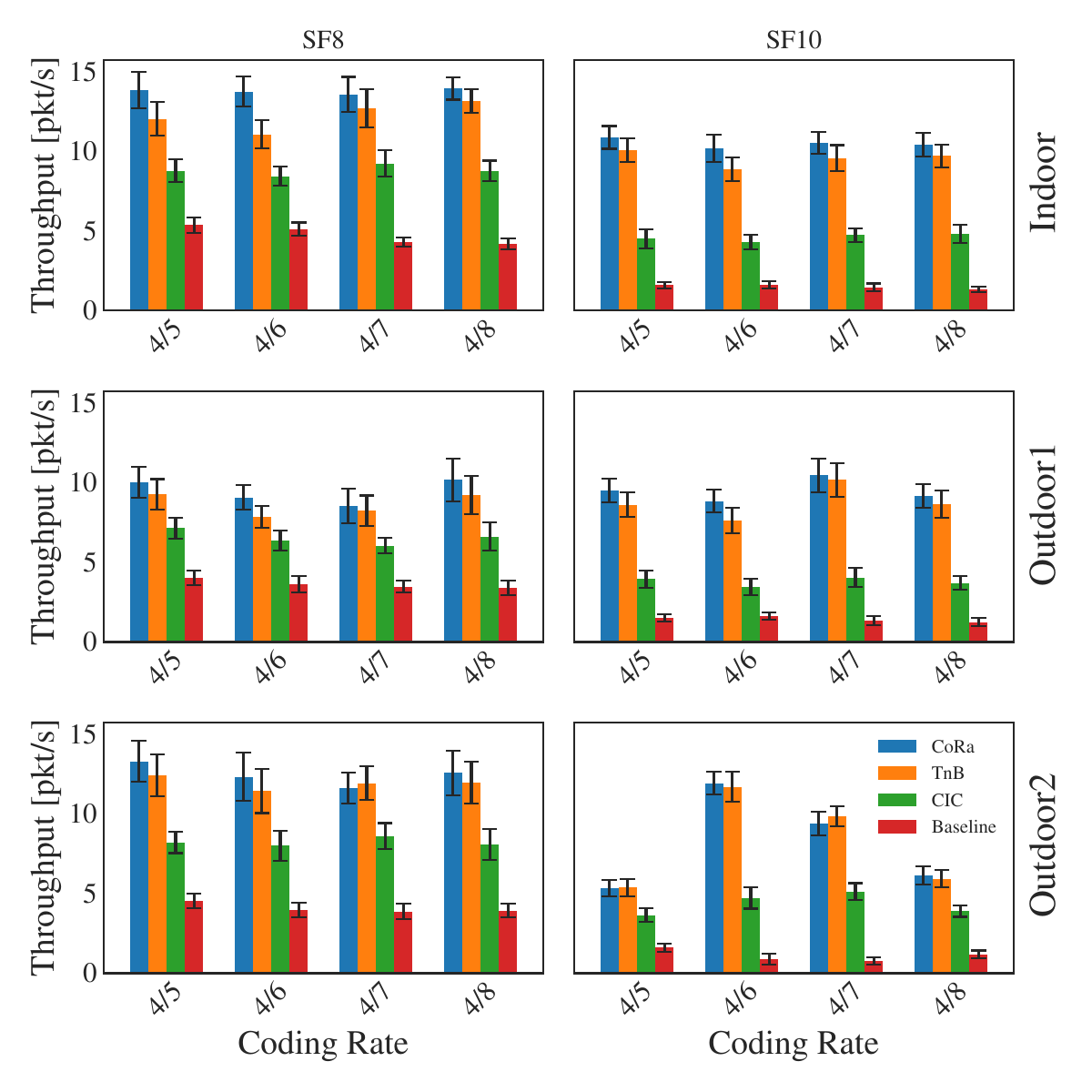}
    \caption{Throughput vs coding rate in the TnB dataset. Devices with spreading factor
	10 and coding rates 4/5 and 4/8 transmit 20 pkt/s, while devices in
	al other scenarios transmit 25 pkt/s.}

\label{fig:tnb_dataset}
\end{figure}

For the TnB dataset (see \autoref{fig:tnb_dataset}), the \our demodulator performs
comparably to or better than TnB. It shows significantly 
enhanced performance  in scenarios with high SNR and lower spreading factor (SF),
achieving up to 24\% higher throughput than TnB (indoor, SF8, CR/4/6).

A decrease in SNR, as experienced in the outdoor scenarios (SF8, left-center
and left-bottom), degrades the estimation of the Bayesian classifier features
due to increased noise. This results in lower throughput and reduces the 
 gap between \our and TnB, as seen in Outdoor1
(15\% improvement over TnB at CR 4/6) and Outdoor2 (7\% improvement over TnB at
CR 4/6). For CR4/7 and CR4/8 in the Outdoor1 and Outdoor2 scenarios (SF8, left),
TnB and our demodulator perform statistically comparable.

In the SF 10 scenarios (see \autoref{fig:tnb_dataset}, right), our solution shows a smaller performance advantage than for the SF8 scenarios,
with an average improvement of 9\% in the Indoor scenario and 8\% in the
Outdoor1 scenario. In the CR4/7 case in Outdoor1 (SF10, right), as well as for all Outdoor2 scenarios (SF10, right), TnB and \our{}
perform statistically comparable.
Note that in the Outdoor2 scenario (SF10, right), the CR4/5 and CR4/8 scenarios have notably less throughput due to the capture scenario with a transmission rate of only 20 packets per second.

These results confirm that while CIC always outperforms the baseline, it still falls short
compare to TnB and our solution. The baseline achieves a maximum of 5 packets per second in the SF8 scenario (indoor, CR4/5) and up to 1.5 packets per second in the SF10 scenario (outdoor1, CR4/6).
In contrast, our approach exhibits an average throughput improvement of $2.2\times$ compared
to the CIC demodulator and $7.4\times$ compared to the Baseline.

\subsection{Resistance to Multipath Fading and Doppler Effect}

\begin{figure}
    \includegraphics[width=0.5\textwidth]{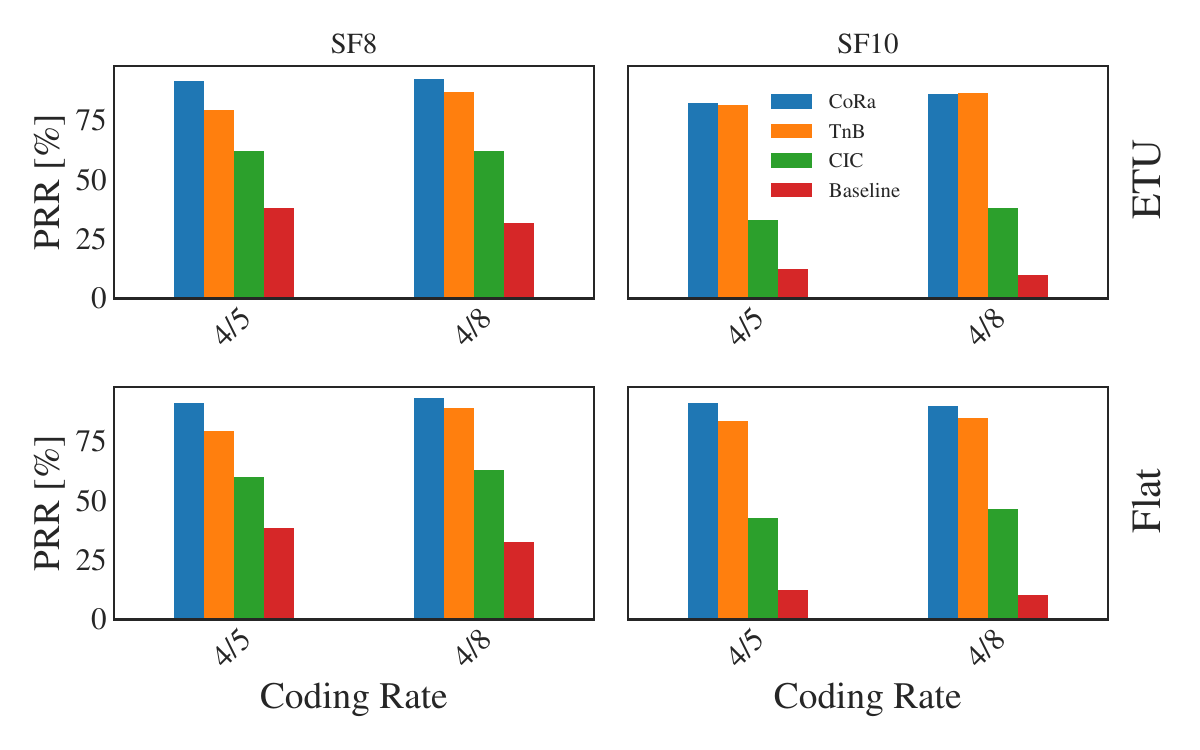}
	\caption{Packet Reception Ratio vs. coding rate in the simulated dataset for LTE Extended Typical Urban (ETU) and Flat channels. Devices transmit at 15 pkt/s with SF8 (SNR: $[-5, 15]$ dB) and SF10 (SNR: $[-10, 10]$ dB), using a 125 kHz bandwidth.}

\label{fig:sim_dataset}
\end{figure}

The real-world captures of the CIC and TnB datasets do not account for mobility nor environments notably affected by multi-path fading, which are common in dense urban
areas. To evaluate the robustness of our solution in such conditions,
we continue with  simulated packet traces from an LTE Extended
Typical Urban (ETU) channel model~\cite{3gpp-tr36873}. This model is characterized by strong
fluctuations and multi-path fading. Specifically, the channel is defined by a Power Delay
Profile (PDP) with 9 taps, each undergoing Rayleigh fading. The channel
also features delay spread of $5 \mu s$ and a maximum Doppler shift of $ 5 Hz$.

We generate simulated captures of 900 LoRa transmissions using \textit{NumPy}. These
simulations include scenarios with both the ETU channel and a flat channel, the latter
of which does not incorporate multi-path fading or Doppler shift.

 \autoref{fig:sim_dataset} presents the packet reception ratio for the evaluated
demodulators.
In the SF8 scenario under the ETU channel conditions (top-left),
our demodulator outperforms the current state-of-the art solutions, despite the strong
 channel fluctuations. Specifically, our demodulator improves
the packet reception ratio over TnB by
approximately 12.3\% with a coding rate of 4/5 and by about 5\% with a coding rate of 4/8. In the SF10 scenario (top-right in the figure),
both the \our and the TnB demodulators perform similarly, with less than one
percent  difference.

When comparing the variations of packet reception ratios  between the ETU and flat
channels, we observe that the TnB algorithm and the baseline are more robust
against channel fluctuations. The TnB decoder varies only around 2\% 
 between ETU and flat channels, while the baseline shows a variation of about one percent. In
contrast, the CIC and the \our demodulator are more affected by strong
fluctuations, as illustrated in the SF10 scenarios (\autoref{fig:sim_dataset}, top-right and bottom-right).
Specifically, in the ETU channel, the packet reception ratio for CIC and \our demodulator decreases by 9 resp. 8 percentage points, with
a coding rate of 4/5, and by 8 resp. 4 points at the coding rate 4/8.

\subsection{Performance Overhead}\label{sec:overhead}

\begin{figure}
    \includegraphics[width=0.5\textwidth]{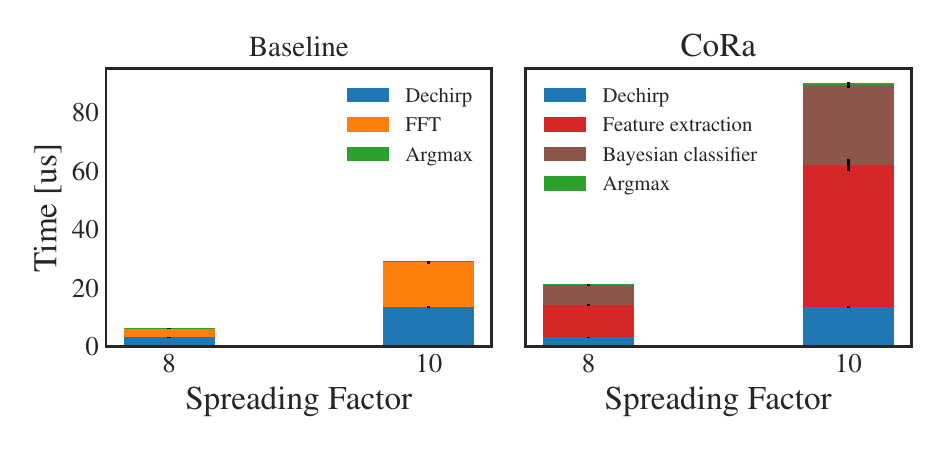}
    \caption{Execution time per symbol for standard LoRa (Baseline) and \our demodulators at SF8 and SF10, showing stage-wise contribution to total demodulation time.} 
\label{fig:complexity}
\end{figure}

We observe from \autoref{fig:complexity} that the \our demodulator spends approximately 50\% of the time on
feature extraction, where the primary load is the calculation of two Fast
Fourier Transforms (FFTs). This operation is computationally efficient with a
complexity of $O(N log N )$. Approximately 30\% of the time is spent on the
Bayesian classifier, which includes the index
calculation ($O(N)$) and the evaluation in the grid ($O(1)$).
The dechirping procedure, which includes applying the carrier frequency offset corrections from the synchronization
algorithm, accounts for around 15\% of the time. The Argmax operation ($O(N)$), which identifies the bin with the highest probability of being the true peak, is negligible in terms of computational time.

Thus, the overall complexity of our demodulator is $O(N log N )$, primarily
dictated by the feature extraction stage.
In comparison to the equally complex Baseline demodulator we observe  a constant overhead factor of approximately 3 -- regardless of the spreading factor.

\section{Discussion}\label{sec:discussion}

\paragraphS{Which are the main advantages of the \our demodulator?}
The evaluation results provide compelling evidence that the \our demodulator is
capable of successfully demodulating frames, even under conditions of severe
collisions and significant channel variations, as well as in mobile scenarios.

The \our symbol detector offers the following advantages over existing work.
First, it does not rely on peak detection, which is prone to errors, particularly
in environments with low Signal-to-Noise Ratio (SNR). Furthermore, our
demodulator operates independently of symbol boundary information. This characteristic
enables the demodulation of symbols without the need to wait for the frame
synchronization of the next incoming frame. In related work, a symbol is impacted by an
incoming frame, because of which the demodulator may be unable to determine whether it is a colliding frame
unless it waits for detecting this next symbol, which requires tracking up to
12.25 symbols (comprising 8 preamble symbols, 2 Start Frame Delimiter (SFD) symbols and 2.25 downchirp symbols).
This  not only increases latency, as the demodulator cannot immediately
classify the symbol without the necessary symbol boundary information, but also
escalates implementation complexity by necessitating the monitoring of pending symbols and awaiting the subsequent frame.

Moreover, the independence from symbol boundary information enhances the versatility 
of our symbol detector across various scenarios. One such scenario is Frequency Hop
Spreading Spectrum (FHSS), where the LoRa transceiver transmit packet fragments across multiple channels.
This feature is standard in LoRa devices and should not be confused with LR-FHSS,
which employs GMSK modulation and narrowband signals.
Another scenario involves accurately detecting symbols even if the frame synchronization
mechanism fails to detect colliding frames.
In case of FHSS, when the frequency hopping pattern is unknown -- such as when
independent LoRa networks transmit data simultaneously -- it becomes impossible to 
predict potential symbol collisions since symbols may appear on different
channels. This limitation renders certain algorithms ineffective, such as TnB or CoLoRa,  
which are designed to match peaks to transmitting devices.

\paragraphS{Can \our Process Packets in Real-time?}
The analysis presented in \autoref{sec:overhead} indicates that the \our demodulator
exhibits a constant overhead of approximately 3 times over the baseline LoRa demodulator.
The computational complexity of our solution is $O (N log N)$, dominated by the complexity of the Fast Fourier Transform (FFT).
Consider for example a Digital Signal Processor (DSP) operating at 200\,MHz, which requires 25
cycles per point for computation of the FFT. In a scenario with $N=256$ samples (spreading factor 8, bandwidth 125\,KHz) the time required for the FFT is calculated as follows:
$$
T_{256}= \frac{25 \cdot 256 \cdot {\mu}s}{200,000,000} = 32{\mu}s
$$
Given that the symbol time in this scenario is $T_s=1.024$\,ms, and assuming the demodulator spends half of the time on FFTs (see \autoref{fig:complexity}),
the total time spend on demodulation is approximately 64\,$\mu$s. This duration represents roughly 6\% of the total symbol time. Hence, the \our complexity in packet processing does not challenge real-time.

Our demodulator exhibits significantly lower computational complexity than the CIC
symbol detector, which requires $(1 + 2 K)$ Fast Fourier Transforms (FFTs) per colliding symbol,
where $K$ is the number of detected colliding frames.
Hence, the complexity of CIC increases polynomially with
the number of collisions, whereas our solution remains constant.

The Thrive (TnB) algorithm involves three main computational steps: FFT
calculation ($O(N log N)$), peak detection ($O(N)$), and peak cost evaluations
($O(N)$). These peak cost evaluations are performed up to $2 M^2$ times per
symbol, where $M$ denotes the number of colliding packets within the
symbol. Since $M \ll N$, where N denotes the number of samples per symbol, the overall
complexity of Thrive simplifies to $O(N log N)$, aligning with our solution,
where the FFT step dominates the total computational cost. Still, the runtime
of TnB cost evaluations scales quadratically, leading to unbounded execution
time. In our solution, the runtime remains constant.

\section{Conclusions and outlook}\label{sec:conclusions}

In this study, we introduced the \our symbol detector, engineered to accurately identify LoRa
symbols even under conditions of severe collisions. Our methodology 
utilized a Bayesian classifier to discern the true symbol peak amidst the peaks from colliding symbols
within the spectrum of the dechirped signal. The foundation of \our lies in harnessing
the symmetry properties of a complete waveform, identified as the true peak,
and the observation that the magnitude of the true peak remains relatively constant within a frame.
This approach facilitates the calculation of robust features, ultimately
enabling the precise identification of the true peak within the dechirped
symbol.

We replaced the symbol detector of a state-of-the-art method (TnB) to develop a
robust LoRa demodulator. This demodulator underwent rigorous evaluations with 
comparison against recent state-of-the-art demodulators and the baseline LoRa
demodulator. The assessment utilized both real-world and simulated data to
comprehensively evaluate its efficacy and performance.

Results indicated that our approach yields outcomes that are either superior or at least
comparable with those of the state-of-the-art solutions, even under conditions of
sub-noise Signal-to-Noise Ratio (SNR). The performance advantage of \our is particularly
evident in scenarios characterized by extreme collisions. Notably, our solution
maintains a constant overhead, specifically three times the demodulation times of the baseline LoRa demodulator.
This suggests the potential feasibility of the implementation on real hardware.

The present study opens two future research directions. First, further research
should refine the feature calculation to account for channel
fluctuations, which holds promise for boosting symbol detection performance in
scenarios with strong multipath fading or Doppler effect. Second, implementing and
deploying \our on a Software Defined Radio (SDR) will facilitate a
comprehensive evaluation in more complex real-world scenarios.

\paragraph{Artifacts} The authors have provided public access to their code and data at \url{https://doi.org/10.5281/zenodo.14582713}

\pagebreak

\bibliographystyle{IEEEtran}
\bibliography{own,rfcs,ids,ngi,iot,layer2,meta,complexity,internet,theory,programming}

\end{document}